\documentclass[%
 reprint,
superscriptaddress,
 amsmath,amssymb,
 aps,
]{revtex4-2}

\usepackage{comment}
\usepackage{graphicx}
\usepackage{dcolumn}
\usepackage{bm}
\usepackage{xcolor}
\usepackage{siunitx}
\usepackage{soul}

\newcommand{\bo}{\boldsymbol}

\newcommand{\mean}[1]{\left\langle #1\right\rangle}
\newcommand{\br}[1]{\left( #1\right)} 

\begin{document}

\author{John Geiger}
\affiliation{Fachbereich Chemie, Universit\"at Konstanz, 78457 Konstanz, Germany}
\author{Niklas Grimm}
\affiliation{Fachbereich Physik, Universit\"at Konstanz, 78457 Konstanz, Germany}
\author{Matthias Fuchs}
\affiliation{Fachbereich Physik, Universit\"at Konstanz, 78457 Konstanz, Germany}
\author{Andreas Zumbusch}
\affiliation{Fachbereich Chemie, Universit\"at Konstanz, 78457 Konstanz, Germany}

\preprint{APS/123-QED}

\title{Decoupling of rotation and translation at the colloidal glass transition}

\date{\today}

\begin{abstract}
Little is known about the coupling of rotation and translation in dense systems. Here, we report results of confocal fluorescence microscopy where simultaneous recording of translational and rotational particle trajectories from a bidisperse colloidal dispersion is achieved by spiking the samples with rotational probe particles. The latter consist of colloidal particles containing two fluorescently labelled cores suited for tracking the particle's orientation. A comparison of the experimental data with event driven Brownian simulations gives insight into the system's structure and dynamics close to the glass transition and sheds new light onto the translation-rotation coupling. The data show that with increasing volume fractions, translational dynamics slows down drastically, whereas rotational dynamics changes very little. We find convincing agreement between simulation and experiments, even though the simulations neglect far-field hydrodynamic interactions. An additional analysis of the glass transition following mode coupling theory works well for the structural dynamics but indicates a decoupling of the diffusion of the smaller particle species. The shear stress correlations do not decorrelate in the simulated glass states and are not affected by rotational motion.
\end{abstract}

\maketitle

\section{\label{sec:level1} Introduction }

At the colloidal glass transition, the flow properties of a dispersion change at the colloid  scale \cite{Hunter2012}. A single colloid in a continuum fluid exhibits Brownian random motion, and the diffusion coefficients of translation and rotation $D_t$ and $D_r$, respectively, are coupled according to the laws of Stokes, Einstein, Debye, and Perrin \cite{Dhont,Konderink2003}. The solvent flow has to obey the condition of vanishing relative velocity on the surface of the colloid. Thus, random forces and torques are coupled leading to a constant ratio $D_t/D_r^l=d^2/3$ for a sphere of diameter $d$, independent e.g.~of the viscosity $\eta$ of the dispersion. In the semi-dilute regime, the coupling of rotation and translation of colloids has been a topic of longstanding interest, especially the question whether the solvent or the dispersion viscosity set the rotational diffusion coefficient \cite{Jones1989,Felderhof1993,Degiorgio1994,Lettinga2000,Poblete2014}. At higher concentration, the close relationship between rotation and translation gets lost when approaching solidification. The direction of the deviations depends on the type of liquid, the thermodynamic phases explored, and the kind of tracer particle studied  \cite{Roos2016,Hsiao2017,Ilhan2022,Chun2021,Schuetter2017,Chong2009,Vivek2017}. In glass-forming liquids, the sensitivity of the rotation-translation coupling has been used to inspect local motion, which remains incompletely understood when approaching the glass transition. Large tracers in molecular fluids have shown a strong variation of the ratio of diffusivities in experiments and simulations: while rotational diffusion remains strongly coupled to viscosity, $D_r \propto 1/\eta$, translational diffusion decouples \cite{Fujara1992,Cicerone1995}. Investigations of colloidal suspensions promise to give further insight into these phenomena since individual particles can be tracked even at concentrations close to the glass transition \cite{Weeks2000}. Direct observation of their rotational motion, however, has been possible only in two systems. Edmond et al.~considered tetrahedral clusters of colloids in a hard sphere solution \cite{Edmond2012}. They observed slowing of translational diffusion slows down less than rotational, that remains tied to structural relaxation. This is reminiscent of the molecular glass-forming liquids. A quite different behavior was observed with spherical Janus colloids \cite{Anthony2006,Kim2011}, where translational diffusion was found to slow down more strongly than rotational when the colloidal glass transition is approached. Clearly, the question of rotation-translation coupling at the colloidal glass transition remains intriguing.

Here, we report results of experiments and simulations of the translation-rotation coupling of dispersions of spherical colloids near the colloidal glass transition and compare them to predictions from mode coupling theory (MCT) \cite{Goetze}. Spherical colloids arguably are the simplest system as they can be described by the paradigmatic hard sphere fluid model. Crystallization can be prevented by 
mixing two species with appropriate size and concentration ratios. The dominant control parameter is the total volume fraction $\phi$. Experiments were based on confocal fluorescence microscopy of dense dispersions of polymethylmetacrylate (PMMA) core/shell colloids with hard potentials. These dispersions were spiked with low concentrations of spherical PMMA colloids containing two cores as rotational probe particles \cite{Schuetter2017}. From the fluorescence data, particle trajectories could be obtained simultaneously for translational and rotational diffusion.  For the simulations, a rough sphere model was employd that proved useful for understanding the collective angular motion in molecular fluids, and had been studied by theoretical and numerical methods \cite{Chandler1974a,Chandler1974b,Chandler1975,Lyklema1979,Berne1977,Pangali1977}. It presumes that the spheres collide elastically and the relative velocity of the contact points reverts in the collision. Because there is no slip at the collision, the angular velocities of the collision partners get changed according to the conservation of total angular momentum. Varying the the mass density distribution, the model has e.g.~provided insight into the origin of different coefficients in the Stokes-Einstein relation in liquids  \cite{Kravchenko2011}, and it is widely used in supercritical fluids \cite{Medina2012} This model of rough spheres  can be tuned from the molecular scale, where atoms move ballistically, to the mesoscopic scale, where colloids perform Brownian motion. The model enables to separate the importance of interparticle correlations from the random forces and torques exerted by the solvent. Thus the cooperativity of the translational and rotational motion can be tested. 
Recently, long-ranged stress correlations in glass-forming colloidal dispersions have been predicted but not yet been observed \cite{Maier2017,Vogel2019}. Employing the Brownian simulations after calibration to the experimental system, we explore the magnitude of the stress correlations in order to address the possibility of their measurement in glass-forming colloidal dispersions.
 
We not only approach the colloidal glass transition but also enter the glass state where aging indicates that the system has fallen out of equilibrium. Quantitative Brownian dynamics simulations accompanying the experiments verify the aging phenomena, verifying that a colloidal glass was prepared. The recorded mean square displacements indicate that large and small spheres are tightly localized, yet there still remains local free volume enabling fast translational and rotational motion. In the simulations, we can tune the Brownian noise and thereby see that angle variations get completely decorrelated by the random torques in a thermal heat bath, while molecular systems require many particle collisions before angular diffusion becomes effective. Finally, we also compare results from experiment and simulation to mode coupling theory (MCT) \cite{Goetze}, that predicts a glass transition value for our mixture which is close to the observed one \cite{Voigtmann_2011}.  Within MCT, rotational motion of anisotropic particles was also considered, predicting a peculiar decoupling of even and odd angular momenta states for small aspect ratios \cite{Schilling1997,Franosch1997,Chong2005}, which however lies at finite aspect ratio and thus cannot be accessed with spherical particles.

The paper is organized as follows. In Sec.~\ref{sec:setups} the system is introduced together with the experimental and simulational methods. In Sec.~\ref{sec:results}, the results are presented including radial distribution functions (RDF), mean squared displacements (MSD) and angular correlation functions for three different densities. Experimental data are analysed and compared to data from the simulations. Sec.~\ref{sect:mct} compares with MCT predictions and analyses the emergence of elasticity. The conclusions and a discussion of the results are given in Sec.~\ref{sec:discussion}. 

\section{Setups} \label{sec:setups}
\subsection{Experiment}
The system under experimental investigation is based on a binary mixture of strerically stabilized PMMA/PMMA core/shell particles that were prepared following the procedure of Schütter et al.\cite{Schuetter2017}. In brief, we first synthesized PMMA core particles with a diameter of $1~mu m$. The cores were copolymerized with either the Bodipy dye (4,4-difluoro-8-(4-methacrylatephenyl)-3,5-bis-(4-methoxyphenyl)-4- bora-3a,4a-diaza-s-indacene)\cite{Baruah2005} or Quasar670. Several seeded dispersion polymerization steps were employed to grow additional PMMA layers onto the cores to obtain particles with a final diameter of 2.08$\mu m$ (polydispersity index (PDI) of 5.1\% determined from electron microscopy images) and of 4.02$\mu m$ (PDI of 3.5\%). Particles with two cores were synthesized by treating the particles with acetone \cite{Schuetter2017}. Two-core particles with a diameter of 4.06$\mu m$ were employed as rotation probes. Several sedimentation steps were performed to reduce the number of particles with more than two cores as far as possible. For the preparation of the colloidal dispersions, the particles were dissolved in a density and refractive index matching mixture of cyclohexylbromide saturated with tetra-butyl-ammonium bromide and cis-decaline (85\% : 15 \%, w/w) \cite{Yethiraj2003}. The number ratio of small particles, larger particles, and rotation probes was 100:100:1. The reported packing fraction measure the total volume taken by all colloidal particles. Steinhardt order parameters were determined to verify that colloidal glasses were prepared \cite{Steinhardt1983}. 

\begin{figure}[ht]
    \centering
    \includegraphics[width=\linewidth]{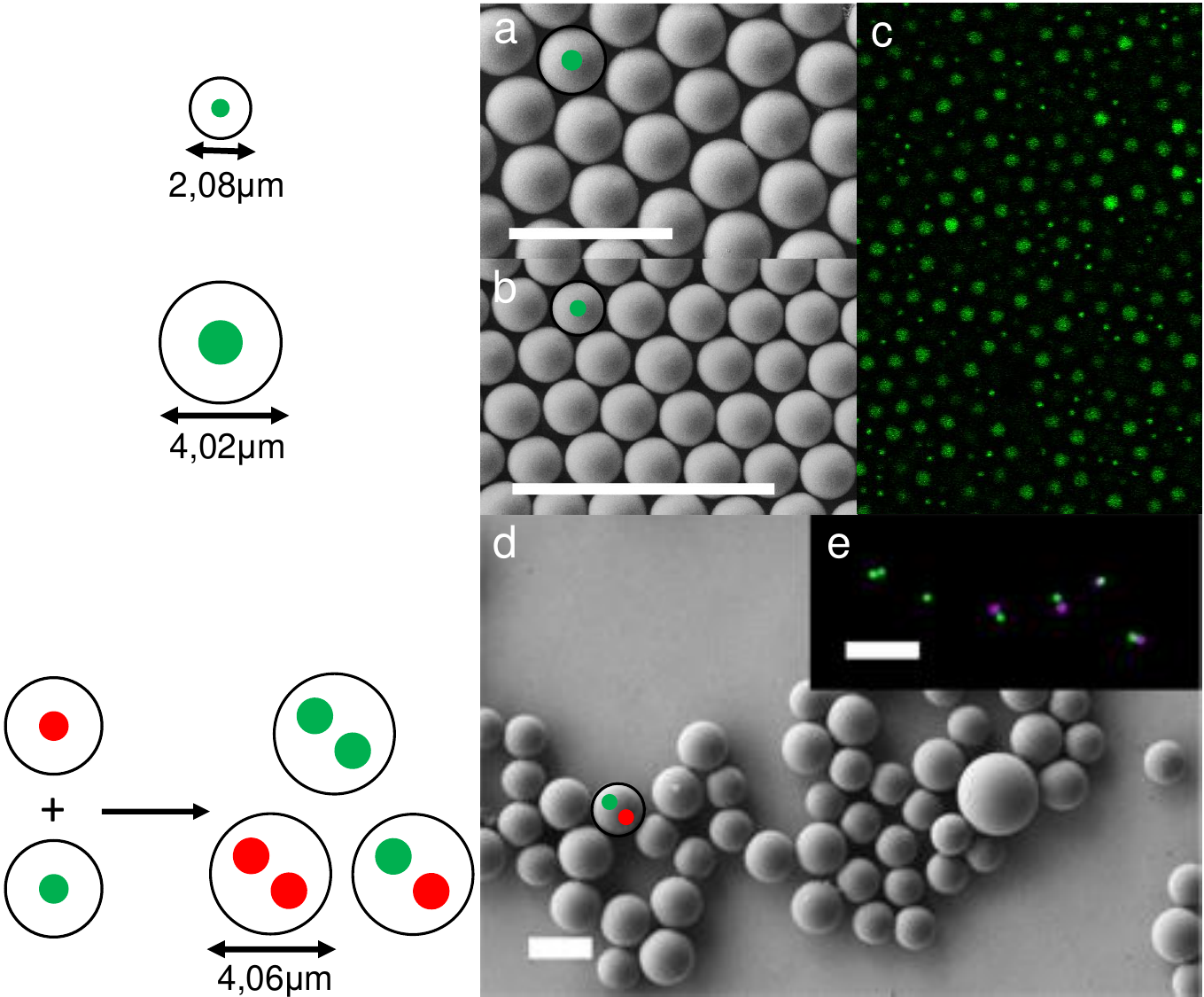}
    \caption{The colloid dispersions consisted of PMMA/PMMA core/shell particles with differently labelled, fluorescent cores. To avoid crystallization, bi-disperse mixtures of particles with diameters of 2.08$\mu m$ and 4.02$\mu m$ were used as a host system. Colloids with two cores and a diameter comparable to the larger host particles were added in high dilution and served as rotation probes.}
    \label{fig:particles}
\end{figure}

Fluorescence images were acquired using a confocal fluorescence laser scanning microscope (TCS SP5, Leica Microsystems) with a resonant scanner (8000\,Hz, bidirectional scanning mode) and a glycerol immersion objective (63x magnification, 1.3 NA). The microscope was temperature stabilized (Ludin Cube 2, Life Imaging Services) to 23\,$\pm$\,0.05$^{\circ}$\,C. 3D-image stacks (1024x256x100 voxels) with pixel sizes of dx=dy=141.3\,nm and dz=210\,nm resulting in 3D image volumes of 144.7x36.2x21.0\,$\mu$m$^3$ were recorded with lag times of t=4\,s. The image stacks were recorded with distances of at least 25\,$\mu$m to the sample chamber walls. To allow equilibration, all samples were put onto the objective 16 hours before the measurement starts. From the image stacks, trajectories for 3D-translation and rotation of the colloids were generated as described earlier \cite{Roller2018}. 

\subsection{Simulation}

In order to interpret and analyze the experimental observations, we performed event driven Brownian dynamics (EBD) simulations of a binary hard sphere mixture \cite{Scala2007,Romano2011}. Different from previous simulations of colloids which neglected their rotational motion, we account for the coupling of rotation and translation via the flow of the solvent. In the EBD algorithm, the particles move with constant velocities $\pmb{v}_i$ and angular velocities $\pmb{\omega}_i$ between events according to Newton's first law. Diffusion is modeled by the Brownian thermostat, redistributing the particles' (angular) velocities at fixed 'Brownian' time steps. Particle collisions encode the potential interactions.

Solvent-mediated interactions among the colloids, called hydrodynamic interactions, are instantaneous and affect the short-time diffusion of the particles \cite{Dhont}. The EBD algorithm  neglects their long-ranged contribution as do other potential free algorithms of hard colloids \cite{Heyes1993}. For smooth spheres, this approximation has been tested and verified when approaching the colloidal glass transition \cite{Marenne2017}. Here, we account for the aspect of the local flow of solvent around the colloids to couple rotation and translation. We do this by two strategies, first by including the correct rotational and translational diffusion in the dilute limit, and, second, by collisions that account for forces and torques.

To model diffusion, we draw the component of the translational velocities in each spatial direction from a Gaussian distribution with zero mean and unit variance $v_0^2=1$ at fixed times steps $\tau_t$ (in Newtonian units this  Maxwell-Boltzmann distribution corresponds to $v_0^2=k_BT/m=1$). Rotational diffusion is achieved by redrawing angular velocities every time step $\tau_r$. Assuming a Maxwell-Boltzmann distribution of angular velocities for spheres with a moment of inertia $I_\zeta = \frac{1}{\alpha}  m d^2_\zeta$,  -- where $\zeta \in \{s,l\}$ denotes either small $s$ or large $l$ particles, $d_\zeta$ the respective diameter, and $\alpha$ is a dimensionless number that encodes  the mass distribution of the spheres --, the  angular variance becomes $\omega_{0,\zeta}^2=\alpha/d_\zeta^2$ for the small and large particles. We assume a homogeneous mass distribution,  which gives $ \alpha = 10$. After many Brownian time steps, the mean squared translational displacements and angular displacements grow linearly with the number of steps and the (short-time) translational and rotational diffusion coefficients take the values $D_t=v_0^2 \tau_t/2$ and $D_r=\omega_0^2 \tau_r/2$. In order to recover the ratio of rotational and translational diffusion coefficients predicted by Stokes, Einstein, and Debye for dilute colloidal spheres \cite{Dhont}, $D_t/D_r^l=d_l^2/3$, 
the ratio between rotational and translational Brownian time steps thus needs to be chosen as \cite{Romano2011} 
\begin{gather}
\tau_t/\tau_r= \alpha /3. \label{eq:ratio}
\end{gather}

At finite density, the HS interact as perfectly rough elastic spheres \cite{Chandler1974a,Chandler1974b,Chandler1975,Lyklema1979,Berne1977,Pangali1977}. This collision rule couples rotations and translations. It inverts the relative velocities of the contact points at collision $\bo{v}_c'=-\bo{v}_c$, as would hold for two colliding rubber balls. There is no slip of the two colliding surfaces. Before the collision, the  contact points approach each other with velocity:
$\pmb{v}_c = \pmb{v}_2 -\pmb{v}_1 +  ( \frac{d_1}{2} \pmb{\omega}_1 + \frac{d_2}{2} \pmb{\omega}_2)\times \hat{\pmb{n}} $ with the collision vector $\hat{\pmb{n}}= (\pmb{r}_1-\pmb{r_2})/|\pmb{r}_1-\pmb{r_2}|$. The relative velocity of the contact points after the collision, $\pmb{v}_c'$, follows from the final velocities (marked by a ') given in the collision law:
\begin{gather}
    \pmb{v}_{1,2}' = \pmb{v}_{1,2} \pm  \Delta \pmb{p}/m_{1,2}  \\
    \pmb{\omega}_{1,2}' =  \pmb{\omega}_{1,2} - \frac{d_{1,2}}{2I_{1,2}} \hat{\pmb{n}}\times  \Delta \pmb{p}\\
    \Delta \pmb {p} =  \frac{2 \mu}{1+\mu \kappa} \left(  \pmb{v}_c  + \mu \kappa  (\pmb{v}_c \cdot \hat{\pmb{n}}) \hat{\pmb{n}}   \right)   
\end{gather}
We introduced the two convenient parameters, reduced mass $\mu=\frac{m_1\,m_2}{m_1+m_2}$, and reduced moment of inertia $\kappa =\frac{\alpha_1}{4 m_1 }+\frac{\alpha_2}{4 m_2 }$, which capture the collisions of homogeneous spheres with different radii. 

The shown time dependent data are averaged over roughly 100 systems, with particle numbers of $N=1728$ up to $N=8\,000$. The number ratio is $N_s/N_l=1$ and the aspect ratio is $d_l=2 d_s$, as determined in the experiment. Only the density is varied. In order to sample the dynamics, the Brownian timestep of translation is set to $\tau_t \, D_t/d_l^2 = 5\times10^{-7}$. It has to be chosen small enough, so that the observation of short time diffusion in the MSD is possible before the caging regime appears. This is hard for the used packing fractions up to $\phi=0.65$, why we had to use such a small value for $\tau_t$. 

\subsection{Equilibration and simulation preparation}

As we study samples above the glass transition,  the structural relaxation time is excessively long and aging affects the measurements. Thus the protocol to prepare the simulated systems needs to be specified. 

The particles start as point particles from a simple cubic lattice and with initial velocities drawn from the described Gaussians. Then they are propagated by Newtonian dynamics, i.e.~there is no Brownian time step. Every fixed time step $\tau_i=0.05\,v_0/d_l$ all particles are inflated until they are in contact with their closest neighbor or they reached their maximum size \cite{Stillinger1990}. Then we proceeded with Newtonian dynamics and measured the decay of the coherent density correlation function at a wave vector magnitude of the position of the nearest neighbor peak in the static structure factor. This was possible for all shown densities, except for the highest two. The latter cases were stopped earlier because the waiting times increased drastically. We explicitly checked that for these two volume fractions the data shown in Sec. \ref{sec:results} were stationary in the given time window. The waiting time $t_w=0$ is measured as time span since the instance when all particles reach their prescribed radii and with that the prescribed density of the system is reached. The inflation process needs an increasing time-span for increasing densities. For $\phi=0.59$ this process finishes on average after $4.8  v_0 / d_l$ and for $\phi=0.65$ after $18.3\,v_0/d_l$.

\section{Results} \label{sec:results}
We present experimental data compared to simulation for three different densities: below, closely above and clearly above the predicted MCT glass transition \cite{Voigtmann_2011}. For the given size ratio and concentrations, microscopic MCT predicts $\phi_c^{\rm MCT}=0.52$. Considering the tendency of MCT to  overestimate the glass-formation \cite{vanMegen1991,Nauroth1997}, we estimate the experimental value to lie 15\% higher, $\phi_c^{\rm exp}\approx 0.6$, and test it below. We compare the structure by presenting the RDF, the translational dynamics by measuring the MSD, and the rotational dynamics by showing averaged Legendre polynomials of the angular variation.

\subsection{Crystallization}

The phase diagram of the binary mixture studied in this work is investigated in Ref.~\onlinecite{Engel2020}. The authors report the appearance of $AlB_2$ crystalline states for several densities. Within the density range explored in this work, crystalline states are reported for $\phi=0.60$ and $\phi=0.62$. For the experimental systems, Steinhardt order parameters \cite{Steinhardt1983} of 4th and 6th order were determined and showed no sign of crystalline order. To test for the possible formation of crystals in the simulated systems, we calculated bond order parameters of the final configurations. The coarse grained bond order parameters are introduced in \cite{Dellago2008}. To calculate the nearest neighbors we follow the recipe in \cite{vanMeel2012}. Our systems do not show signs of crystallization, neither the experimental samples nor the simulated systems. We attribute this to the difficult pathway to Laves-phase formation observed and overcome in Refs.~\cite{Engel2020,Schaertl2018}.

\subsection{Radial distribution functions}

\begin{figure}[ht]
    \centering
    \includegraphics{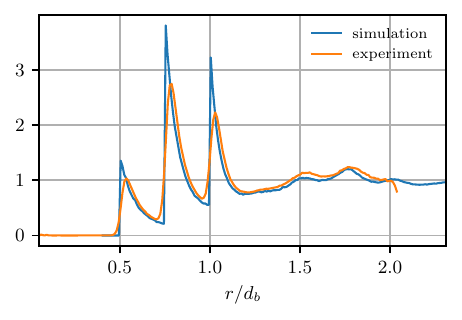}
    \caption{Radial distribution function of the systems with experimental density $\phi_{exp}= \num{0.574} $ and a density in simulation of $\phi_{sim}=0.59$. }
    \label{fig:rdf_1}
\end{figure}

\begin{figure}[ht]
    \centering
    \includegraphics{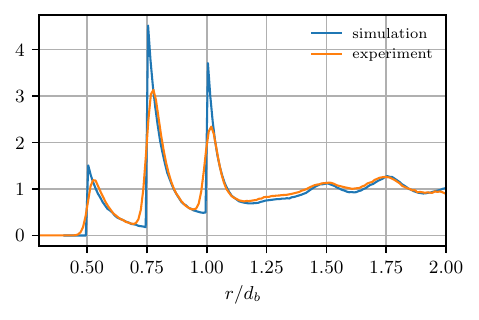}
    \caption{Radial distribution function of the systems with $\phi_{exp}= \num{0.634} $ and $\phi_{sim}=0.62$.  }
    \label{fig:rdf_2}
\end{figure}

\begin{figure}[ht]
    \centering
    \includegraphics{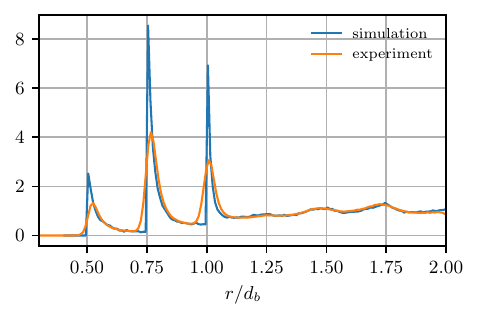}
    \caption{Radial distribution function of the systems with $\phi_{exp}= \num{0.658} $ and $\phi_{sim}=0.65$.  }
    \label{fig:rdf_3}
\end{figure}

Figures \ref{fig:rdf_1}, \ref{fig:rdf_2} and \ref{fig:rdf_3} show the RDF results for increasing densities. Three next-neighbour peaks are visible indicating the three distances of small-small, small-large, and large-large spheres. Weak traces of these distances can also been seen at the second neighbour shells. For increasing total packing fraction, the structuring in the RDF gets stronger, but no sharp peaks registering crystalline order emerge. All states are fluid like.  

We use the RDF to calibrate the simulated systems to the experimental ones, which suffers from the following problems: $(i)$ The assumption of a perfectly binary mixture does not hold for the experiment, where a small polydispersity exists. $(ii)$ The determination of packing fractions of colloidal dispersions is error prone due \cite{Poon2012}. The first problem is overcome by choosing a single length scale, the diameter of the large species $d_l$, when comparing simulation and experiment. The experimental RDF data originally measured in $\SI{}{\micro \metre}$ are scaled by the value $d_l  = \SI{4.76}{\micro\metre}$, which provides the visual best fit. The simulated packing fractions, chosen for the fits, then turn out very close to the experimentally determined values. We thus consider the differences between simulation and experiment to arise from polydispersity and from the problem to measure $\phi$ precisely. 

Figure \ref{fig:rdf_1} shows the RDF for simulation and experiment in the liquid phase. The simulation graph has sharp edges at distances $d_s$, $1/2\,(d_s+d_l)$ and $d_l$ resulting from the exact hard sphere interactions and the exact binary particle diameter distribution. Compared to that, the experimental graph is smoother, pointing towards the (small) polydispersity in the experimental sample.

Figure \ref{fig:rdf_2} shows the comparison with the identical value of $d_l$ at the higher packing fraction around the MCT glass transition, whereas Fig.~ \ref{fig:rdf_3} shows it deep in the glass state. 
Also for these higher volume fractions, values of $\phi_{sim}$ can be chosen in the simulations that are rather close to the experimentally predetermined values. 

Overall, a very good mapping between the structure as captured in the two-component radial distribution functions is observed between simulation and experiment. As seen in many investigations of glass-forming systems, the structure changes only smoothly when crossing the colloidal glass transition \cite{Hunter2012}.

\subsection{Mean squared displacements}
\label{sec:MSDs}
\begin{figure}[ht]
    \centering
    \includegraphics{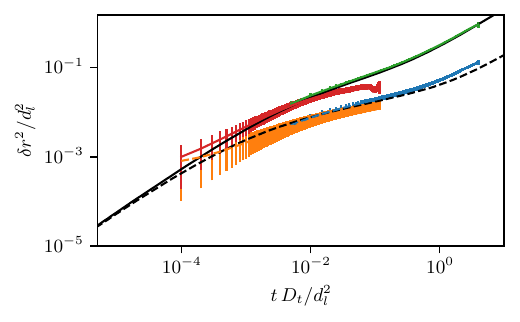}
    \caption{MSD of small (upper lines) and big particles (lower lines) in experiment (colors) and simulation (black). Experimental data at short times are from 2D confocal scans, while 3D scans gave the data at long times. The volume fraction of the experiment is $\phi_{exp}= \num{0.574} $, while the simulation was performed at $\phi_{sim}=0.59$ }
    \label{fig:msd_charge1}
\end{figure}

\begin{figure}[ht]
    \centering
    \includegraphics{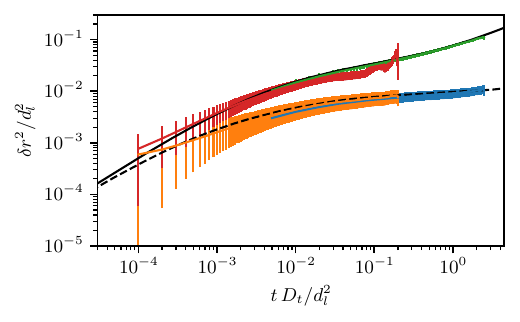}
    \caption{MSD of experiment and simulation with volume fractions $\phi_{exp}= \num{0.634} $ and $\phi_{sim}=0.62$. Simulation was started after propagating the system for $t_w=19\,000 d_l/v_0$ in Newtonian dynamics. }
    \label{fig:msd_charge2}
\end{figure}

\begin{figure}[ht]
    \centering
    \includegraphics{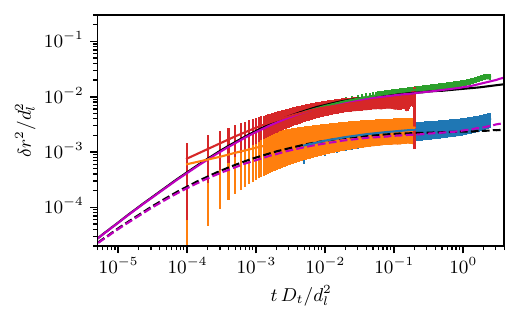}
    \caption{MSD of experiment and simulation. With volume fractions $\phi_{exp}= \num{0.658}$ and $\phi_{sim}=0.65$. The magenta line shows data of a system that was given less time for equilibration. Hint to aging.  }
    \label{fig:msd_charge3}
\end{figure}

After having studied the structure, we now turn to the dynamics of the system. 
Figures~\ref{fig:msd_charge1}, \ref{fig:msd_charge2}, and \ref{fig:msd_charge3} depict the MSDs $\delta r^2_\alpha = \mean{(\pmb r_\alpha \br{t} - \pmb r_\alpha\br{0})^2}$, where $\alpha$ denotes the particle species, and colored symbols indicate the experimental data. The systems are the ones whose structure was studied above. The higher slopes are MSDs from small host particles, whereas the lower ones originate from the large host particles. The experimental short time data were determined from 2D measurements, while data for longer times were extracted from 3D measurements. The two measurements are denoted by different colors. Simulation curves are drawn in black.
 
 While the previous section showed that the structure of the binary mixture changed smoothly with increasing packing fraction, the MSD data highlight that the colloid motion slows down strongly \cite{Williams2001,Hunter2012}. This is caused by the trapping of the colloids in the cages of their neighbours for intermediate times, which can be recognized from the emerging plateau in the MSD. The finite lifetime  plateau indicates incipient localization, and a localization length following Lindemann's discussion \cite{Barrat1989} of crystal melting could be read-off. Yet, determining the long-time diffusivity is difficult because of the finite time-window of observation, and because of possible aging effects. Here, the simulations can help to interpret the dynamics, because the waiting time can be controlled, and additional states can be sampled.
 
Since the particle diameter was taken from the RDF comparison, the MSD can be plotted in re-scaled form 
$\delta r_\alpha^2/d_l^2$ with
$d_l^2=(\SI{4.76}{\micro\metre})^2$.
As the packing fractions has also been determined in the fits of the RDF, the only parameter to match the time-dependent data from simulation and experiment is the time scale $d_l^2/D_t$, given by the translational short-time diffusion coefficient.  Best agreement was achieved by re-scaling experimental data by 
\begin{gather}
    D_t^{exp}/d_l^2 = \SI{0.001}{\sec^{-1}} \label{eq:difftExp}. 
\end{gather}
Figures \ref{fig:msd_charge1} shows the diffusion in a colloidal fluid state. Long-time diffusion of both particle species is clearly attained at long times. At intermediate times, a plateau in the MSD starts developing which indicates the incipient localization of the colloids. As expected, the larger colloids move less than the smaller ones, and the gap widens with increasing time. 

At the intermediate packing fraction $\phi_{exp}= \num{0.634} $, the MSD have slowed down, and long-time diffusion of the large colloids cannot be seen in the experimental window anymore. Yet, the apparent diffusion of the smaller colloids at long times suggests that the state could be fluid.  This is, however, not supported by the simulations at this packing fraction. The system at $\phi_{sim}=0.62$ cannot be equilibrated, and exhibits a structural relaxation time that grows with waiting time $t_w$. The MSD in the window shown in Fig.~\ref{fig:msd_charge2}, however, are already stationary and do not change with increasing $t_w$.

Figure \ref{fig:msd_charge3} shows a density far above the predicted MCT-glass transition. Therefore aging effects would not come as a surprise, and aging effects can indeed be observed in the simulations, shown by the magenta line. We simulated a system that was equilibrated by using the standard inflation protocol and then followed by Newtonian dynamics for $t_w=1000\,d_l / v_0$. This reproduces the earlier onset of the $\alpha$-relaxation in the experiment very well. For longer waiting times, however, the escape out of the cage takes longer.  In Fig.~\ref{fig:msd_charge3}, the black curve shows the MSD of a system that aged for $t_w=16\,000\,d_l/v_0$, whereas for larger $t_w$ no ageing effects remain in the studied time window. 

Concluding the comparison of the translational dynamics, one can summarize that very good agreement can be achieved by mere tuning of the short-time diffusion coefficient and transferring all other parameters from the structural comparison.

\subsection{Rotational dynamics}

\begin{figure}[ht]
    \centering
    \includegraphics{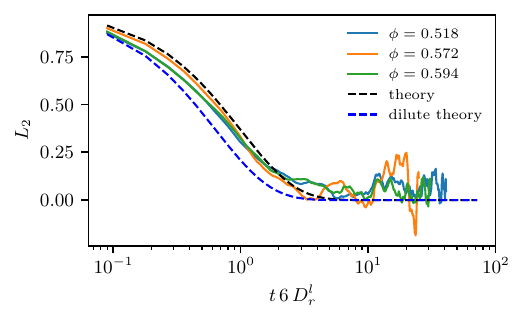}
    \caption{ 
    Experimental angular correlation function $L_2$ for different densities. Theory of the free Brownian rotation Eq. \eqref{eq:freeRot}, where $D_r$ is given in Eq. \eqref{eq:rotDiffExp}.  The density dependency is very weak, especially compared to the extreme sensitivity of the MSDs in this density range.   }
    \label{fig:rotExpSimC2}
\end{figure}

\begin{figure}[ht]
    \centering
    \includegraphics{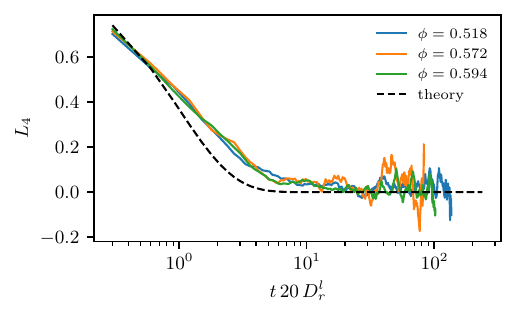}
    \caption{Experimental angular correlation function $L_4$ for different densities. Theory of the free Brownian rotation Eq.~\eqref{eq:freeRot}, where $D_r$ is given in Eq.~\eqref{eq:rotDiffExp}.}
    \label{fig:rotExpSimC4}
\end{figure}

To quantify the rotational dynamics we calculate Legendre polynomials $L_n$ of the angle variation of the tracer. The second polynomial, which can also be measured by depolarized light scattering \cite{Hoffmann2009}, reads
\begin{gather}
    L_2(t) = \mean{ \frac{3}{2} (\cos \left( \theta(t) \right)^2 -\frac{1}{2} } \label{eq:L2} ,
\end{gather}
where we chose $\theta(0) = 0$ for each particle.
The idealized case of diffusion on a sphere yields \cite{Berne}   
\begin{gather}
    L_n(t) = \exp\br{-n\,(n+1) \,D_r \,t }  \label{eq:freeRot}.
\end{gather}
Experiment and free diffusion are compared in Fig.~\ref{fig:rotExpSimC2}. 
The ratio of the rotational and translational diffusion coefficients of a single spherical colloid of diameter $d_l$ is given by the Stokes-Einstein-Debye and the Stokes-Einstein relation to $D_t/D_r^l=d_l^2/3$. From Eq.~\eqref{eq:difftExp} we can then predict the free rotation coefficient of the experiment to read
\begin{gather}
    D_r^{l} = \SI{0.003}{\sec^{-1}} \label{eq:rotDiffExp}
\end{gather}
The black dashed theory slope in Figs.~\ref{fig:rotExpSimC2} and \ref{fig:rotExpSimC4} is calculated with Eqs.~\eqref{eq:freeRot} and \eqref{eq:rotDiffExp}. From this it can be concluded that rotational diffusion close to the glass transition is still very similar to rotational diffusion of a dilute sphere. Figure \ref{fig:rotExpSimC4} shows the fourth Legendre polynomial which is affected by smaller angle variations and thus records somewhat larger differences to Debye's free rotational diffusion.  

Considering the strong slowing down of the translational motion observed in Figs.~\ref{fig:msd_charge1} to \ref{fig:msd_charge3}, the findings in Figs.  \ref{fig:rotExpSimC2} and \ref{fig:rotExpSimC4} state  that the rotational diffusion of a sphere is not affected by the packing fraction. This holds even in the extreme case when crossing the colloidal glass transition. In the limit $\phi \to 0 $, the rotational diffusion is that of a free rotor, and it remains that of a free rotor at any measured density. There is a slight subtlety, though, arising from the solvent mediated hydrodynamic interactions which make $D_t$ and $D_r$ slightly density dependent. The blue dashed line in Fig.~\ref{fig:rotExpSimC2} is calculated by Eq. \eqref{eq:freeRot} with $D_r^l$ directly taken from measurements of dilute tracer particles. It does not agree as well with the measured data because the increased particle density has changed the local friction experienced by the colloids. This is captured in the small density-dependence of the short-time rotational diffusion constant, while the long-time diffusion constant does not change relative to this short-time one. 

\begin{figure}[ht]
    \centering
    \includegraphics{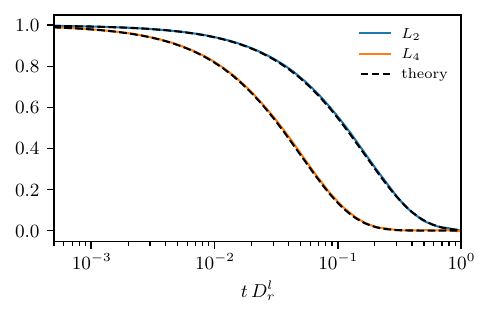}
    \caption{Simulation results for the rotational correlation functions Eqs.\,\eqref{eq:L2} of the large particles. Dashed lines show the theory of the free rotation \eqref{eq:freeRot}.
    \label{fig:rotSim}
    }
\end{figure}

Simulation results at $\phi=0.65$
are shown in Fig.~\ref{fig:rotSim}. We see perfect agreement between the theory of free Brownian rotation and simulation. This can be rationalized because we choose $\tau_t$ smaller than the mean collision time to produce proper short time diffusion. From Eq. \eqref{eq:ratio} it follows, that $\tau_r$ is even smaller. So the rotational Brownian events of the thermostat are much more frequent than collisions with other particles. The mean collision time at $\phi=0.65$ in BD is $\tau_c=40\cdot 10^{-7}\,D_t/d_l^2$, which means a rotational termostat event is roughly 26 times more frequent then a collision. This implies, that the thermostat wipes out the effect of the collisions on the rotational diffusion. 

To this point one could argue that this is the same for the translations as well, because translational velocities are also resampled eight times more often than collisions happen on average. Yet, this clarifies the essence of the cooperative cage effect: The translation of a spherical tracer is hindered by the surrounding particles, and diffusion requires a collective rearrangement. This leads to long-lasting memory in the particle collisions that are not destroyed by random velocity changes. By contrast, the rotation of a sphere does not require extra volume (in contrast to, for instance, tetrahedra \cite{Edmond2012} or ellipsoids \cite{Roller2021}, or particles at interfaces \cite{Niggel2023}). Thus, the rotation of a sphere does not couple to the cage of the surrounding particles and no (strong) excluded volume effects are observed for the diffusion.

\section{MCT glass transition analysis}
\label{sect:mct}

In order to estimate the glass transition point, we performed an $\alpha$-scaling analysis following the predictions of  MCT. It predicts a scaling of the density auto correlation (ACF) function, $S(q,t)=\mean{\rho\br{\pmb q,t}^* \rho\br{\pmb q,0}}$, during the decay from the plateau ($\alpha$-relaxation) for packing fractions approaching the MCT glass transition $\phi^{MCT}_c$ \cite{Goetze}. Here,  $\rho\br{\pmb q,t}=N^{-1/2}\sum_{j=1}^N\exp\br{i \pmb q \cdot \pmb r_j(t) }$ is the fluctuating microscopic density.  

\begin{figure}[ht]
    \centering
    \includegraphics{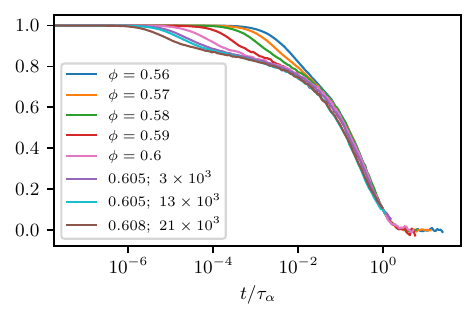}
    \caption{Normalized collective density ACF at $q\,d_l=7.9$. Re-scaled, so that a unit time marks a decay to 0.1. The graphs collaps almost perfectly. The $\alpha$-relaxation times extracted from this figure are shown in Fig. \ref{fig:relaxationTimes}. 
    }
    \label{fig:densityScaling}
\end{figure}

\begin{figure}[ht]
    \centering
    \includegraphics{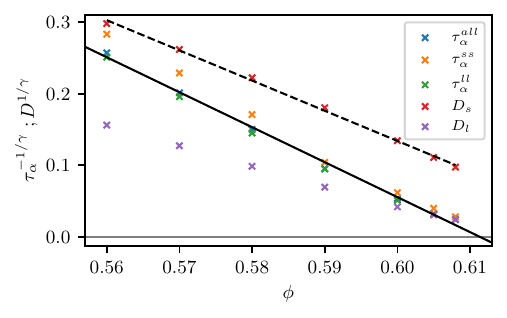}
    \caption{The relaxation times of the total and partial density correlation functions from figure \ref{fig:densityScaling}. By plotting $\tau_\alpha^{-1/\gamma}$ the algebraic law from Eq. \eqref{eq:MCTalpha} is given by a linear graph. The linear fit to the relaxation times predicts a glass transition of $\phi^G_{MCT}=0.61$. A similar fit on $D_s^{1/\gamma}$ predicts a significantly higher divergence at $\phi=0.63$.  }
    \label{fig:relaxationTimes}
\end{figure}

Figure \ref{fig:densityScaling} shows the re-scaled density ACF for different densities; for efficiency the Brownian thermostat is turned off in these simulation runs. The predicted $\alpha$-scaling is well confirmed by the simulations. MCT predicts an algebraic divergence of the relaxation times $\tau_\alpha(\phi)$ towards the glass transition  
\begin{gather}
    \tau_\alpha = C\cdot |\phi - \phi_c|^{-\gamma} \label{eq:MCTalpha}
\end{gather}
with an MCT-exponent $\gamma = 2.6477$ depending on the parameters of the binary mixture such as number and aspect ratios. The quoted value corresponds to size ratio $d_l/d_s=2$ and equal concentrations $N_l/N_s=1$ \cite{Voigtmann_2011}. As can be seen in Fig.~\ref{fig:relaxationTimes}, the divergence is well confirmed by the relaxation times taken from the total, see Fig.~\ref{fig:densityScaling}, and the species-resolved density ACF. In each case, the wavevector corresponds to the average particle separation. The power-law behavior leads to the estimate $\phi_c^{exp}=0.61$. The MCT analysis in Fig.~\ref{fig:densityScaling} includes the diffusion coefficients obtained from the MSD of small and large spheres. The diffusion of the large spheres is strongly coupled to the structural relaxation, but the smaller particles appear not to arrest with the structure at the same packing fraction. $D_s$ rather appears to vanish at an appreciably higher packing fraction. In Ref.~\cite{Weysser2010}, the relation between the appearance of dynamical heterogeneities and the decoupling of coherent and incoherent time scales is discussed. The data in Fig.~\ref{fig:relaxationTimes} show the decoupling in form of different power-law divergences for $D_\zeta$ and $\tau_\alpha$. This is especially apparent for the small particles, where the collective relaxation time $\tau_\alpha^{ss}$ approaches the same singularity as the one of the big spheres $\tau_{\alpha}^{bb}$ , while $D_s$ vanishes at an appreciably higher packing fraction of $\phi=0.63$. This follows when the black dashed fitted slope in Fig.~\ref{fig:relaxationTimes} is extrapolated to zero.  

\section{Shear modulus}

Recently, the rheology of rough colloids has been discussed in the context of shear thickening caused by frictional contacts \cite{Lin2015,Hsu2018,Schroyen2019,Jamali2019,Pradeep2022}. Thus it is of interest to compare the shear moduli of rough and smooth spheres, which can be done using the standard Brownian dynamics algorithm \cite{Scala2007,Romano2011} and our modified one.
The shear modulus is given by the ACF of the shear stresses $\sigma_{\alpha\beta}(t)$
\begin{gather}
    G(t) = \frac{V}{k_B\,T} \frac{1}{3} \sum_{\alpha<\beta}\mean{\sigma_{\alpha\beta}(t) \sigma_{\alpha\beta}(0) }.
\end{gather}
In the smooth sphere case, the shear stresses come from collisions, which are instantaneous and binary \cite{Alder1957,Lange2009}. For the considered system with rotational degrees of freedom, the stress tensor is in general not symmetric, but can be symmetrized following \cite{MartinParodiPershan}. Then the stress tensor reads 
\begin{gather}
    \sigma_{\alpha\beta}(t)=\frac{1}{\Delta\tau_c V}\sum_{\tau_c }\frac{1}{2}\{  \Delta \pmb{p}_{\alpha} \Delta \pmb{r}_\beta + \Delta \pmb{p}_{\beta} \Delta \pmb{r}_\alpha \}.
\end{gather}
Here $\Delta \tau_c$ denotes the time window, over which we average the collisions. The sum runs over all collisions, happening in the time interval $t\in \left[t-\Delta\tau_c/2, t+\Delta\tau_c/2\right]$. $\Delta\pmb{r}$ denotes the center to center vector of the two colliding particles and $\Delta\pmb{p}$ denotes the transferred momentum.

\begin{figure}[ht]
    \centering
    \includegraphics{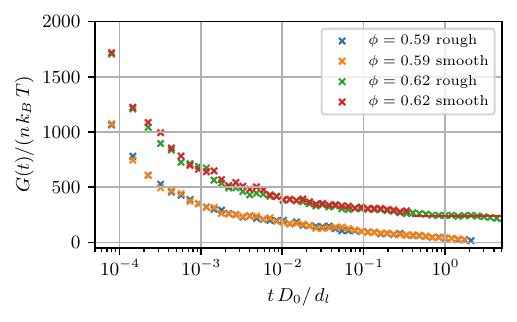}
    \caption{The time-dependent shear modulus $G(t)$  is shown for the binary mixtures of the rough and of the corresponding smooth spheres at a fluid and a glass volume fraction; see the legend. The shear elastic constant of the glass states is marked by a red horizontal bar. \label{fig:modul}  }
\end{figure}

The time-dependent shear modulus $G(t)$ does not show any differences between smooth and rough spheres. For the density above the glass transition, the modulus does not decay in the simulated time window and rather approaches a long time plateau (marked by a horizontal bar in Fig.~\ref{fig:modul}), while it decays to zero in the liquid. The lack of a difference suggests that appreciable asperities are required in order for frictional forces to lock the rotational motion. The amplitude $G_\infty$ of the shear stress correlations frozen-in in the glass state gives the shear modulus \cite{Mewis2012,Fuchs2010} which sets the scale of the long-ranged stress fluctuations in glass-forming dispersions  
\cite{Maier2017,Vogel2019}. It is clearly of entropic origin and scales with the thermal energy per particle, $G_\infty \propto k_BT (N/V)$. Figure~\ref{fig:modul} shows that the prefactor is large compared to unity.


\section{Discussion}
\label{sec:discussion}

We have reported simultaneous measurements of the rotation and translational dynamics of a binary colloidal mixture in the vicinity of its glass transition. The rotations of tailor-made fluorescently labeled multi-core shell tracers could be tracked at very high densities. While the diffusion of large and small particles essentially freezes at the glass transition, the rotations show only a weak dependence on density. This can be rationalized, as the particles are spherical and therefore do not need additional free volume to rotate, while displacements get suppressed by the collective cage effect. 

We performed Brownian event driven simulations modelling the binary mixture precisely, i.e.~taking the size and concentration ratios from experiment. The simulation approach is potential free, viz.~tailored to hard sphere interactions, and mimics local hydrodynamic forces and torques between different particles. The solvent enters the simulation as Brownian thermostat, which leads to the correct rotational and translational diffusion of an isolated colloid, viz.~the Stokes-Einstein-Debye diffusion coefficients. At finite densities, rotation-translation coupling is included in binary collisions, where momentum- and angular momentum-transfer model short-ranged forces and torques. We consider this to mimic lubrication forces transmitted by the solvent. Despite these simplifications, experiment and simulations agree well, considering slow translational diffusion and fast rotational decorrelation. 

For the MSD of the highest density the earlier onset of long time diffusion could be rationalized by aging effects. The simulations show that Brownian fluctuations induced by the solvent, at least as modelled in our scattering law, dominate the initial decorrelation of displacements and angles. However, long-term memory is only encoded in the particle structure not in the solvent flow. The angular motion thus does not couple to the slow cooperative structural rearrangement.

Recently, the emergence of long-ranged stress correlations at the glass transition was predicted \cite{Maier2017}. This requires the macroscopic shear modulus to become slow because it sets the amplitudes of the stress correlations. The simulations indicate that the  shear modulus per particle far exceeds thermal energy close to the glass transition, and that rotational motion has no consequences on the rheology of the investigated dispersions. 

In addition to experiments and simulations, we performed an $\alpha$-scaling analysis of the simulated density correlation function 
that provided an estimate of the MCT glass transition of $\phi_c^{exp}=0.61$, that should be compared to the predicted value  $\phi_c^{MCT}=0.52$. The difference for the considered mixture is similar to the one obtained for slightly polydisperse colloidal hard spheres \cite{Hunter2012}. The algebraic divergence of relaxation times predicted by MCT agrees well with simulations of the local density fluctuations.
Yet, the very different mobilities of the two species cannot be neglected. 
The smaller colloids decouple from the glassy structural relaxation and remain rather mobile around $\phi_c^{exp}$. We also observe stronger ageing of the large particles compared to the small ones. This decoupling was already observed in a unimodal system with broad size distribution \cite{Zaccarelli2015}. In contrast to Ref.~\cite{Zaccarelli2015}, however, the glass transition of the large spheres is not smeared out but remains sharp in our case. 
For even larger size disparity, experiments \cite{Sentjabrskaja2016} and  simulation  \cite{Voigtmann2009} found a qualitative difference in big and small particle motion and anomalous sub-diffusion of the latter.
Even tough we report a decoupling of small and large particles diffusion, we see the same caging and structural arrest for both species.

\begin{acknowledgements}
We thank Thomas Voigtmann for discussions and MCT calculations of binary mixtures. The work was supported  by the Deutsche Forschungsgemeinschaft (DFG) via SFB 1432 project C07. 
\end{acknowledgements}

\bibliography{apssamp}

\end{document}